\documentclass[%
 aip,
 amsmath,amssymb,
 reprint,%
]{revtex4-1}

\usepackage{graphicx}
\usepackage{dcolumn}
\usepackage{bm}

\usepackage[utf8]{inputenc}
\usepackage[T1]{fontenc}
\usepackage{mathptmx}
\usepackage{etoolbox}
\usepackage{xcolor}

\makeatletter
\def\@email#1#2{%
 \endgroup
 \patchcmd{\titleblock@produce}
  {\frontmatter@RRAPformat}
  {\frontmatter@RRAPformat{\produce@RRAP{*#1\href{mailto:#2}{#2}}}\frontmatter@RRAPformat}
  {}{}
}%
\makeatother
\begin{document}

\preprint{AIP/123-QED}

\title{Metal-Insulator transition and Charge Transport Mechanisms in SnSe$_2$ Field-Effect Transistor}
\author{Aarti Lakhara}
 \affiliation{Department of Physics, Indian Institute of Technology Indore, Khandwa Road, Indore, Simrol, 453552, India}
\author{Lars Thole}%
 \affiliation{Institut f\"{u}r Festk\"{o}rperphysik, Leibniz Universit\"{a}t Hannover, 30167 Hannover, Germany}
\author{Rolf J. Haug}
 \email{haug@nano.uni-hannover.de}
 \affiliation{Institut f\"{u}r Festk\"{o}rperphysik, Leibniz Universit\"{a}t Hannover, 30167 Hannover, Germany}
\author{P. A. Bhobe}
 \email{pbhobe@iiti.ac.in}
 \affiliation{Department of Physics, Indian Institute of Technology Indore, Khandwa Road, Indore, Simrol, 453552, India}

\date{\today}

\begin{abstract}

We report an observation of metal-insulator transition in a thin film of SnSe$_2$. The room-temperature carrier concentration of SnSe$_2$ film was increased by electrostatic doping to 1.14$\times$ 10$^{13}$ cm$^{-2}$. A crossover from insulating phase to metallic state was clearly observed. The low-temperature charge transport mechanism is governed by two-dimensional (2D) variable-range hopping. This mechanism is influenced by band bending and gap states introduced by selenium vacancies.  At low temperatures, the mobility is primarily limited by charged impurities, while at higher temperatures, it follows a power-law dependence, $\mu = T^{-\gamma}$, indicating a dominance of electron-phonon scattering. The application of a gate field shifts the Fermi level toward the conduction band, and at sufficiently high temperatures, this drives the system into a metallic state. Our findings offer insights into the charge transport mechanisms in SnSe$_2$ FET, this understanding will allow for the optimization of other 2D materials for advanced electronic device applications.

\end{abstract}

\maketitle
SnSe$_2$, a semiconductor belonging to the group of $s-p$ metal dichalcogenides, has garnered significant attention in recent years due to its exotic quantum phases such as charge density waves\cite{CDW_SnSe2}, and superconductivity\cite{Superconductivty,Superconductivty_2}. Alongside its varied quantum phases, SnSe$_2$ possesses a band gap of approximately 1 eV, which is comparable to silicon (1.1 eV), thereby highlighting its potential for optoelectronic applications\cite{domingo1966fundamental}. Since 2013, SnSe$_2$ has been investigated extensively as a channel material for field-effect transistors (FETs). Some of the recent results include a 25 nm thick device that exhibits a persistent conductive state even under substantial negative back-gate voltages, indicating a high electron concentration \cite{5}. Furthermore, placing a top capping layer of polymer electrolyte on a 10 nm thick FET\cite{8} was seen to help achieve a current on/off ratio of $10^4$. In another study, a 84 nm thick device had been found to demonstrate a high room temperature drive current of 160 $\mu$A/$\mu$m\cite{6}, placing it in competition with other high-performance 2D - based FETs\cite{7}. 

While SnSe$_2$-based FETs have been extensively explored for practical applications, the fundamental physics governing charge transport mechanisms in this material remain unclear. The present study aims at investigating the temperature and electric field-dependent transport characteristics of a 11.8 nm ($\sim$ 19 layer) thick SnSe$_2$ FET fabricated on SiO$_2$/Si substrate through simple exfoliation method. Multi-layer configurations are advantageous as they are less influenced by the substrate compared to monolayers, facilitating extraction of intrinsic material properties in lower dimensions. Mobility serves as a key metric for FET devices, and SnSe$_2$ FETs have shown high mobility, for example, 85 cm$^2$/Vs at room temperature for an 8.6 nm thick device\cite{9}. The present investigation thus covers field effect mobilities ($\mu_{\mathrm{FE}}$) and transfer curves of SnSe$_2$ across a temperature range from 10 K to 300 K. Our back-gated device demonstrates a high mobility of 128 cm$^2$/Vs at 300 K, surpassing previously reported values in the literature. Furthermore, the field-effect mobility increases as the temperature decreases, reaching a maximum of approximately 387 cm$^2$/Vs at 60 K and starts to decrease with further decrease in temperature.  
Our analysis effectively demonstrates that the decrease is due to static charge vacancies present in the system, which are responsible for variable range hopping (VRH) transport in the low temperature range. Interestingly, the VRH insulating state transforms into a metallic state above 120 K, due to increase in carrier concentrations ($>$ 6.66 $\times$ 10$^{12}$ cm$^{-2}$) induced by back-gate voltage.

\begin{figure}
 \centering
    \includegraphics[width=1\linewidth]{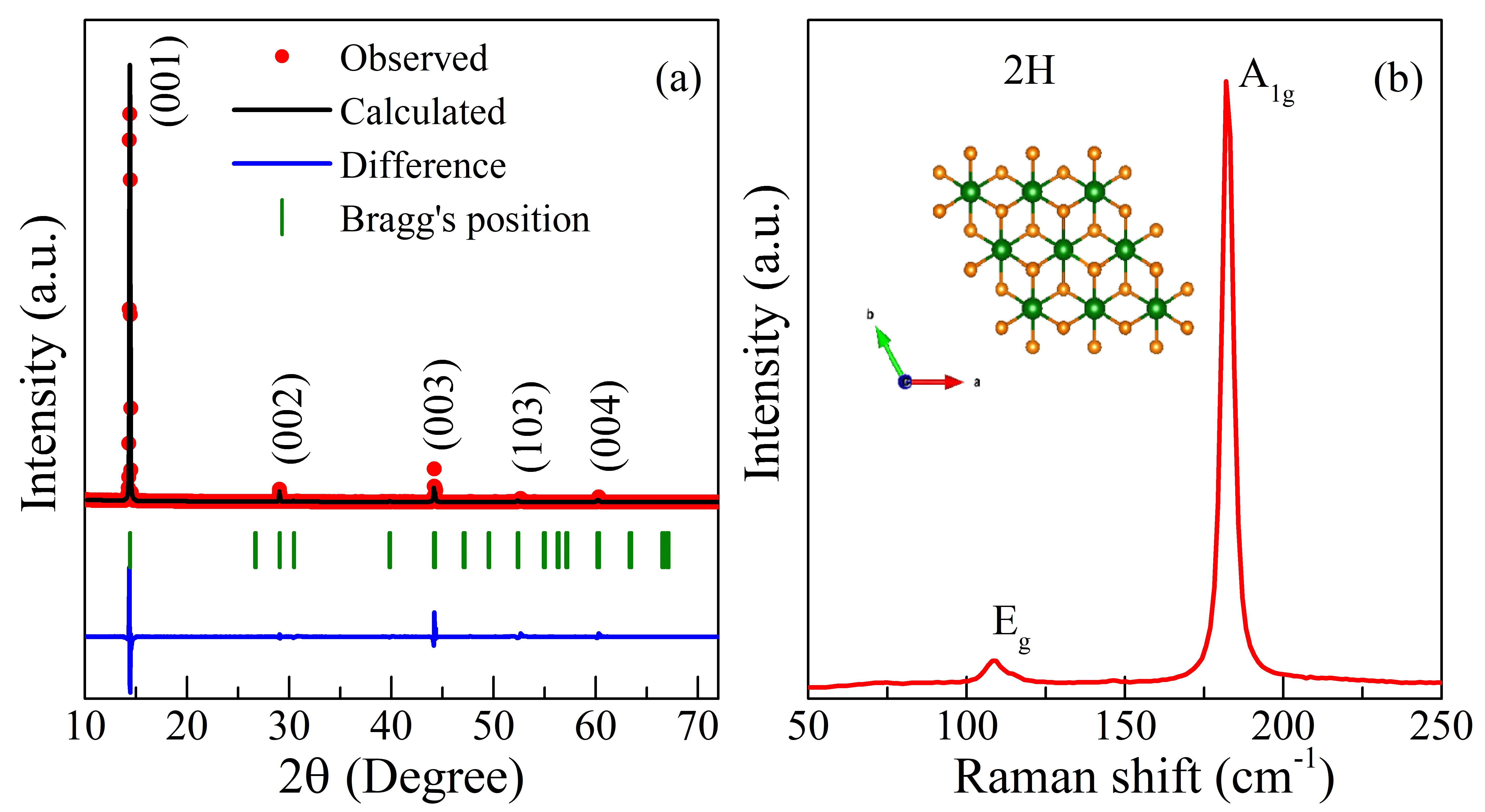}
\caption{\label{xrd}(a) Le Bail refined XRD pattern of SnSe$_2$. (b) Room temperature Raman spectra of SnSe$_2$ single crystal and inset shows 2H structure of SnSe$_2$.}   
\end{figure}
\begin{figure}
 \centering
    \includegraphics[width=1\linewidth]{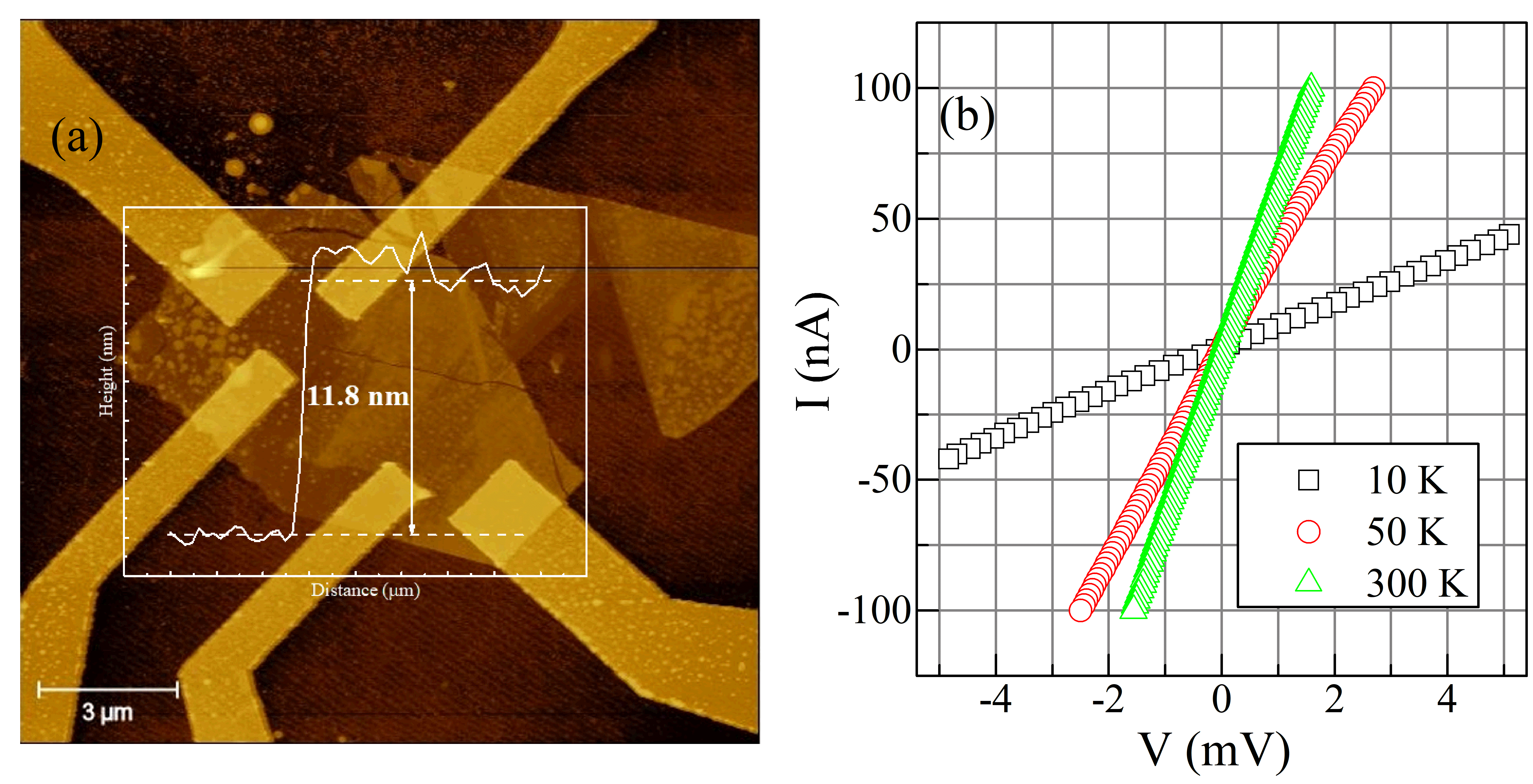}
\caption{\label{AFM} (a) Atomic force microscopy image of SnSe$_2$ thin flake with height profile of the flake, showing a thickness of 11.85 nm. (b) Temperature-dependent I-V curve, demonstrating good electrical contact.}   
\end{figure}

The SnSe$_2$ single crystals were grown by chemical vapor transport (CVT) method (see supplementary material Sec. S1). The elemental mapping and EDX spectra were recorded on bulk crystal to confirm the homogeneity (see supplementary material Sec. S2). An atomic ratio of Sn:Se = 1:1.95 is obtained in synthesized crystals. The XRD profile of  SnSe$_2$, shown in Fig.~\ref{xrd}(a) is an expression of its single crystalline nature that displays highly oriented $(00l)$ planes and very low intensities of other planes. All the observed peaks are indexed to CdI$_2$ type hexagonal unit cell with \textit{P-3m1} space group, confirming the phase purity of the composition. Using the Le Bail method integrated within the FullProf suite\cite{lebail,carvajal}, we carried out the refinement of the XRD profile and extracted the lattice parameters values as, $a=b$ = 3.857 \AA~ and $c$ = 6.147 \AA~, that match well with previous reports\cite{pham2020high}. Fig.~\ref{xrd}(b) shows the Raman spectra of SnSe$_2$, where in-plane vibrational mode ($E_g$), and out-of-plane vibrational mode ($A_{1g}$) are located at 110 $cm^{-1}$ and 185 $cm^{-1}$ respectively. Similar to other dichalcogenides, the Sn atom in SnSe$_2$ is sandwiched between two Se atoms, and it crystallizes in two polytypes: 2H  (space group P$\bar3m$1,  $D^d_{3d}$, 164) and 1T (space group P$6_3/mmc$, $D^4_{6h}, 194$). The peak position of the $E_g$ mode at 110 $cm^{-1}$ confirms the 2H polytype\cite{13, mandal2024low} in the SnSe$_2$ single crystals synthesized here. The SnSe$_2$ FET device was fabricated using standard mechanical exfoliation method (details of device fabrication can be found in supplementary material Sec. S3). The AFM image of fabricated SnSe$_2$ FET is presented in Fig. \ref{AFM}(a).  The ohmic-contact is formed as confirmed through I-V curve measurement (see Fig. \ref{AFM}(b)).

\begin{figure}
 \centering
    \includegraphics[width=1\linewidth]{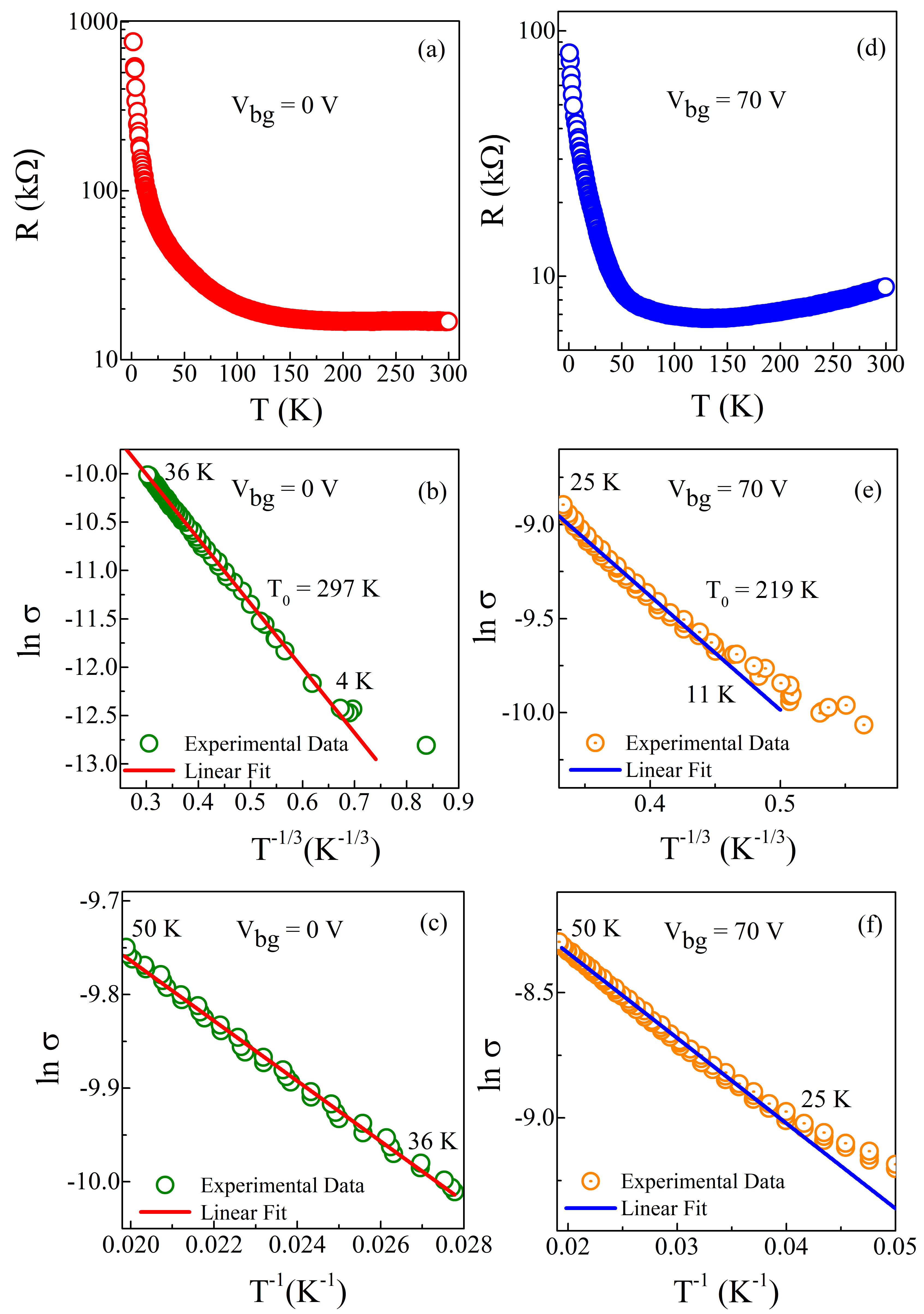}
\caption{\label{vrh}(a) Four-point resistance as a function of temperature at zero back-gate voltage. (b) ln $\sigma$ as function of $T^{-1/3}$. Symbols represent experimental data, and the red line shows a linear fit from 36 K to 5 K, indicating a 2D VRH transport mechanism. (c) ln $\sigma$ as a function of 1/T from 36 K to 50 K, suggesting a transition to nearest-neighbour hopping transport. (d)  Four-point resistance as a function of temperature at a 70 V back-gate voltage. (e) represent the 2D VRH for 70 V back-gate voltage from 25 K to 11 K. (f) From 25 K to 50 K nearest-neighbour hopping describe the transport mechanism very well for 70 V back-gate voltage.}   
\end{figure}

We begin by investigating the characteristics of the electronic states of the SnSe$_2$ channel. For this, we conduct temperature-dependent four-point resistance measurements with the back-gate voltage set to 0 V and 70 V, as shown in Fig.~\ref{vrh}. At 0 V in Fig.~\ref{vrh} (a), SnSe$_2$ exhibits semiconductor behavior, consistent with its intrinsic nature\cite{ying2018unusual, 19}. With increasing temperature the resistance drops sharply until $\sim$ 30 K, followed by a gradual decrease until $\sim$ 150 K, and thereafter remains constant irrespective of the temperature change. 

As seen in Fig.~\ref{vrh}(b), the low temperature data is best described by Variable Range Hopping (VRH) transport mechanism \cite{mott1969conduction}, showing a straight-line dependence between ln($\sigma$) and $T^{-1/3}$, where $\sigma$ is the conductivity. In VRH transport mechanism electrons hop between occupied localized states and free localized states. These localized states are spatially separated but have equivalent energies, and they exist within the band gap of the material. In SnSe$_2$, these "gapped" states can be intrinsic, arising from charge defects in the crystal, or extrinsic, due to the metal-semiconductor interface formed at the electrical contacts. Given the high electron affinity of SnSe$_2$ \cite{highelectron_affinity_SnSe2}, electrons flow from the metal contact (Cr/Au), leading to band bending at the interface and potentially contributing to gapped states. Additionally, recent electronic structure calculations of SnSe$_2$ show that Se vacancies induce several peaks in the gapped region of the density of states, closely linked to the formation of gapped states that activate new transitions in the optical spectrum \cite{zhong2020electronic}. The given Sn:Se ratio of 1:1.95 in the synthesized crystal corresponds to approximately 2.5\% Se vacancies. These vacancies are significant enough to contribute to the formation of impurity levels close to the conduction band,  which in turn lead to the observed variable-range hopping (VRH) conduction at low temperatures. VRH occurs through these states, and is described by the relation $\sigma \propto exp[-(T_0/T)^{1/(d+1)}]$, where $T_0$ is the characteristic temperature and $d$ is the dimensionality of the system. From the fitting, we obtain the characteristic temperature $T_0 \sim$ 297 K. 
Given the 2D nature of SnSe$_2$ crystals, our data fits well for $d$ = 2, indicating that the electrical transport is confined only in the $x-y$ plane. In the intermediate temperature range, where resistance changes gradually, the transport follows the nearest neighbour hopping (NNH) mechanism described by the relation, $\sigma$ $\propto exp(-E_a/k_BT)$, where E$_a$ is the activation energy and $k_B$ is the Boltzmann constant. Accordingly, ln($\sigma$) varies linearly with $1/T$, as shown in Fig.~\ref{vrh}(c).

With the back-gate voltage set at 70 V, the shape of the resistance versus temperature curve remains largely unchanged in the low temperature range. However, the magnitude of resistance decreases significantly compared to 0 V data. Another noticeable change is the increase in resistance with rising temperature, beyond $\sim$ 120 K, a behavior characteristic of metallic conduction, as can be seen in Fig.~\ref{vrh}(d). Accordingly. for temperatures below 25 K, conductivity fits the VRH mechanism, followed by the NNH mechanism until 50 K. Beyond the NNH regime, the effect of rising thermal energy is seen to affect the conduction process. The data here fits an activation transport behavior (as will be discussed further), until metallic state is reached beyond $\sim$ 120 K. 

\begin{figure}
\centering
   \includegraphics[width=0.9\linewidth]{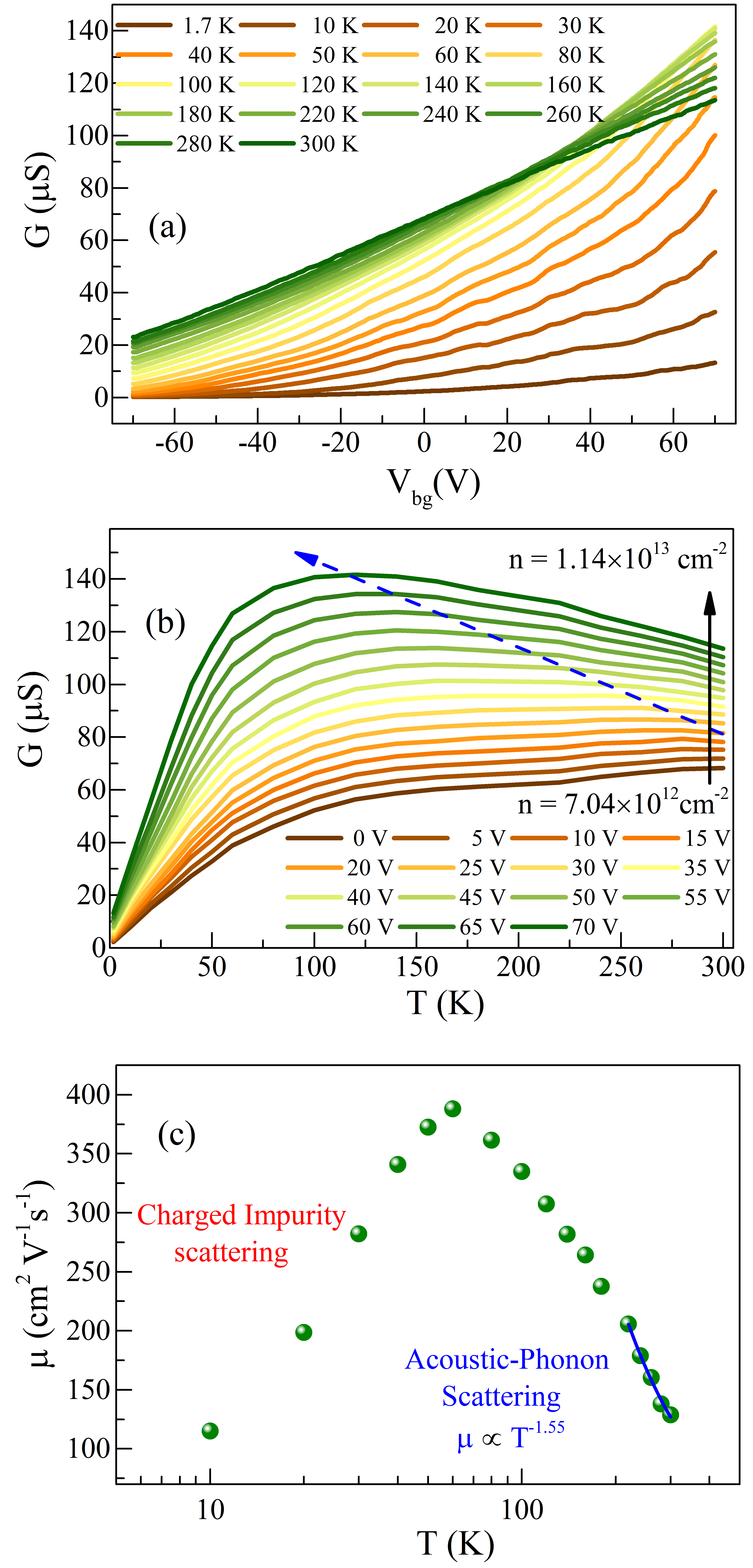}
\caption{\label{cond}(a) Conductance (G) as a function of back-gate voltage (transfer characteristics) at various temperatures. (b) Conductance as a function of temperature for back-gate voltages ranging from 0 V to 70 V. The dashed blue arrow is a guide to eye, showing the metal-insulator transition. (c) Variation of mobility with temperature.} 
\end{figure}

To gain more insight into the metal-insulator transition (MIT), we examine the temperature-dependent conductance behavior with systematic increase in the back-gate voltages. The conductance ($G$) measured by sweeping the back-gate voltage in the range $\mp$ 70 V and at various fixed temperatures, ranging from 1.7 K to 300 K, is shown in Fig.~\ref{cond}(a). As depicted in these plots, conduction predominantly occurs in the positive back-gate region, indicating the intrinsic $n$-type nature of SnSe$_2$\cite{evans1969optical}. The corresponding plot of conductance versus temperature (derived from Fig.~\ref{cond}(a)) at various fixed back-gate voltages, is shown in Fig.~\ref{cond}(b). 

We calculated the 2D carrier density ($n_{2D}$) of SnSe$_2$ as, $n_{2D} = C_{ox} \frac{\Delta V_{bg}}{e}$, where C$_{ox}$ is given by $\epsilon_0 \epsilon_r/d$, where $\epsilon_0$ = $8.85\times 10^{-12}$ F$m^{-1}$, $\epsilon_r = 3.9$ for SiO$_2$, $d$ = 340 nm is the thickness of SiO$_2$ layer, and $\Delta V_{bg} = V_{bg}-V_{th}$ where $V_{th}$ is the threshold value of back-gate voltage beyond which a distinct rise in conductance with increasing back-gate voltage is observed. At room temperature, $n_{2D}$ increases from $7.04\times10^{12}$ cm$^{-2}$ at 0 V to $1.14\times10^{13}$ cm$^{-2}$ at 70 V. The minimum back-gate voltage of 10 V, where the metallic behavior seems to occur at a much higher temperature ($\ge$ 300 K) corresponds to carrier concentration 6.66 $\times$ 10$^{12}$ cm$^{-2}$. This value of $n_{2D}$ is of the same order as that observed in MoS$_2$, where the MIT is reported to occur at $\sim 1\times10^{13}$ cm$^{-2}$ in the presence of a high dielectric environment\cite{18}. As the back-gate voltage is increased, the transition temperature that marks a change in conductance from an insulating type to metallic type, decreases systematically. A dashed line in Fig.~\ref{cond}(b) is a guide to the eye, demonstrating the variation of MIT from 300 K for V$_{bg}$ = 10 to 120 K at V$_{bg}$ = 70 V.

To get further insight into the charge transport mechanism, the field-effect mobility is extracted from the conductance curve in the 50-70 V range of back-gate voltage, using the expression ,$$\mu = \frac{L}{W} \left(\frac{1}{C_{ox}}\right) \frac{dG}{dV_{bg}}$$
 Here L,W are length and width of channel, respectively. As shown in Fig.~\ref{cond}(c), the temperature dependence of mobility exhibits a distinct peak, indicating a characteristic transport mechanism within the system. The maximum mobility value of 387 cm$^2$/Vs is reached at 60 K, and decreases thereafter as temperature increases, reaching a value of 128  cm$^2$/Vs at room temperature. The temperature dependence of mobility indicates an electron-phonon scattering mechanism at play. Indeed, a power-law fit to the mobility data at elevated temperatures, expressed as $\mu = T^{-\gamma}$, yields an exponent $\gamma=1.55$. This value closely aligns with the temperature dependence for acoustic-phonon scattering, $\mu \propto T^{-1.5}$, suggesting it to be the dominant phenomenon at high temperatures. With a decrease in temperature, the mobility values deviate from the power-law fit, indicating that additional scattering mechanisms may be playing a role in the transport properties of SnSe$_2$. 

\begin{figure}
 \centering
    \includegraphics[width=1\linewidth]{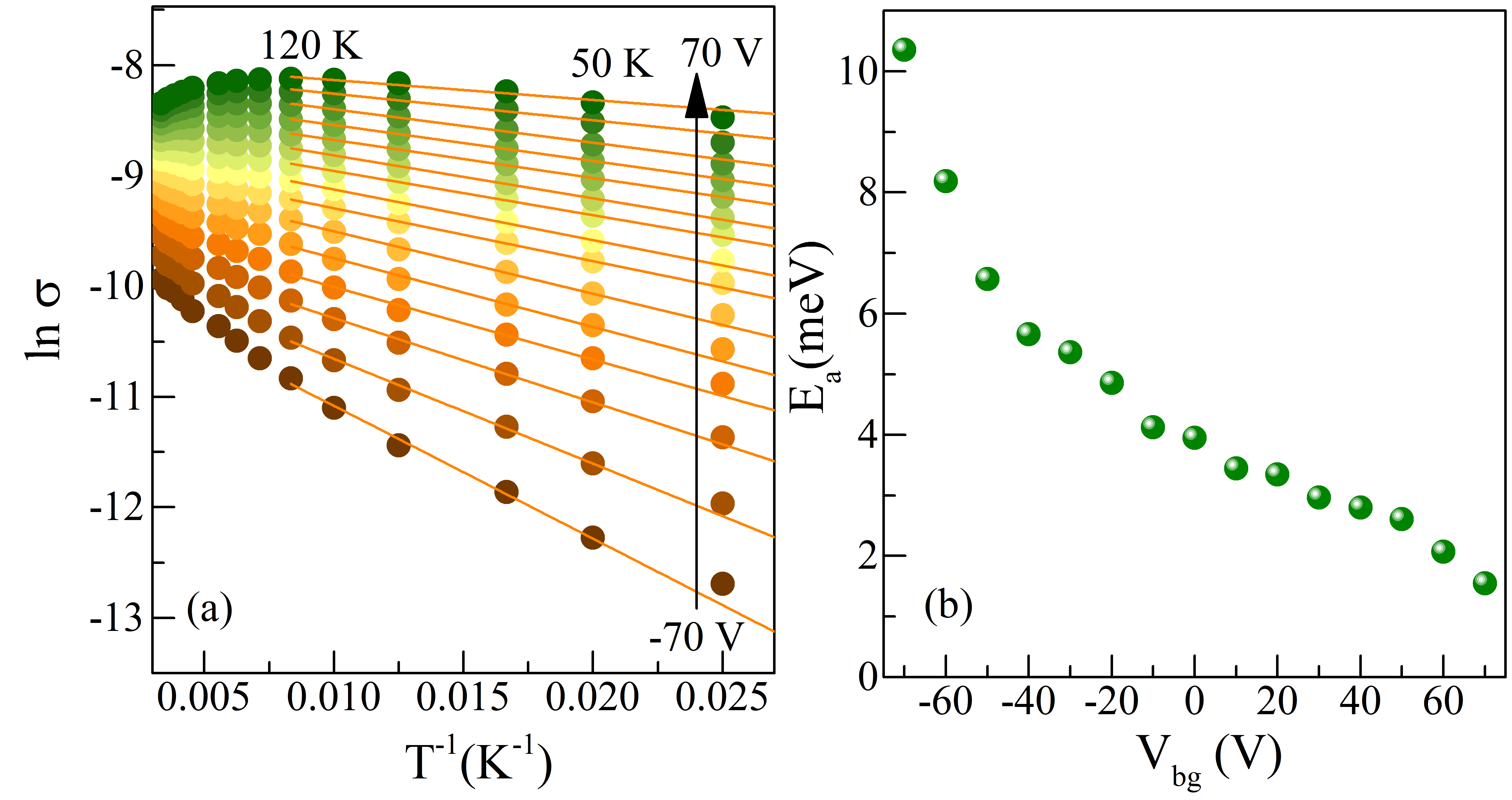}
\caption{\label{actv}(a) A plot demonstrating the thermally activated transport behaviour in conductance measured at various back-gate voltages in the intermediate temperature regime. (b) Variation in the activation energy with back-gate voltage.}   
\end{figure}

Previously, the analysis of the resistance versus temperature curve obtained for the back-gate voltage set at 70 V, indicated that thermally activated transport occurs in the intermediate temperature range. Accordingly, the data extracted from Fig.~\ref{cond}(a), and presented as $ln(\sigma)$ versus temperature in Fig.~\ref{actv}(a), conveys a linear relationship in the temperature range 50 K $\leq$ T $\leq$ 120 K. The extracted values of activation energy (E$_a$) seem to decrease as the back-gate voltage increases (see Fig.~\ref{actv}(b)), indicating a clear influence of the back-gate voltage on the electronic states. As a result of accumulated electrons caused by the field effect, donor energy levels below the conduction band become filled, leading to a shift in the Fermi level. From the plot, it is clear that this shift occurs from 10.35 meV to 1.55 meV. Consequently, SnSe$_2$ become more conductive at 70 V back-gate voltage.

Hence, transition of SnSe$_2$ from an insulating to a metallic state is observed with the application of back-gate voltage. With $\sim$ 2.5\% Se vacancies in the system, their random distribution creates a rough potential landscape for the electrons, contributing to the formation of gapped states between the valance and conduction band edges. At low carrier density, electrons are unable to screen this potential fluctuation, and the transport occurs through the hopping mechanism in the insulating regime. As the carrier density increases through the field effect, electrons can screen the disorder potential more effectively. We  observe that the screening ability is sensitive to temperature. The application of back-gate voltage introduces more electrons into the sample, thereby increasing the carrier concentration. As the back-gate voltage is systematically increased, the Fermi level shifts closer to the conduction band, reducing the thermal energy required to excite charge carriers. This effect is reflected in the observed decrease in activation energy with increasing back-gate voltage.

In conclusion, we conduct a comprehensive investigation of charge transport phenomena in a 19 layers (11.85 nm) thin SnSe$_2$ FET device. We explore the transport characteristics over a broad range of carrier densities, which are tuned by applying back-gate voltages. Notably, we report a metal-insulator transition in SnSe$_2$ at high carrier densities, occurring between 120 K and 300 K. Our analysis indicate that charge impurity scattering dominates the mobility at low temperatures, while phonon scattering becomes the predominant mechanism at higher temperatures in the presence of field effect. Based on our findings, we demonstrate that the phase transition in SnSe$_2$ can be controlled by modulating the carrier density and temperature alone, without the need for complex device architectures. This suggests potential applications where SnSe$_2$ can function as both the semiconducting channel and metallic interconnects, offering a solution to challenges such as band mismatches and Schottky barriers commonly encountered in 2D material-based FETs.

\begin{acknowledgments}
This work was supported by German Academic Exchange Service (DAAD) funded project LUH-IIT Indore Partnership (2019-2023) under the ``A New Passage to India program". Aarti lakhara acknowledges DST-INSPIRE (File: DST/INSPIRE/03/2019/001146/IF190704), New Delhi,
for providing the research fellowship. Work at Leibniz University Hannover was funded by the Deutsche Forschungsgemeinschaft (DFG, German Research Foundation) under Germany's Excellence Strategy - EXC 2123 Quantum Frontiers - 390837967 and within the Priority Program SPP 2244 '2DMP'. We also  acknowledges Department of Science \& Technology, India, for its support through FIST project (SR/FST/PSI-225/2016) for the establishment of Raman spectroscopy measurement system at the Department of Physics, IIT Indore.
\end{acknowledgments}

\section{AUTHOR DECLARATIONS}

Aarti Lakhara and Lars Thole contributed equally to this work.

\subsection{Conflict of Interest}
The authors have no conflicts to disclose.

\subsection{Author Contributions}
Aarti Lakhara: Conceptualization (equal); Data curation (equal); Investigation (equal); Writing –original draft; Writing – review \& editing (equal). 

Lars Thole: Conceptualization (equal); Investigation (equal); Formal analysis (equal); Writing – review \& editing (equal).  

Rolf J. Haug: Conceptualization (equal); Formal analysis (equal); Writing – review \& editing (equal); Resources (equal); Supervision; Funding acquisition (lead).  

P. A. Bhobe: Conceptualization (equal); Formal analysis (equal); Writing – review \& editing (equal); Resources (equal); Supervision; Funding acquisition (lead).

\section*{Data Availability Statement}

The data that support the findings of this study are available from the corresponding authors upon reasonable request.

\nocite{*}
\bibliography{aipsamp}

\providecommand{\noopsort}[1]{}\providecommand{\singleletter}[1]{#1}%
\begin{thebibliography}{21}%
\makeatletter
\providecommand \@ifxundefined [1]{%
 \@ifx{#1\undefined}
}%
\providecommand \@ifnum [1]{%
 \ifnum #1\expandafter \@firstoftwo
 \else \expandafter \@secondoftwo
 \fi
}%
\providecommand \@ifx [1]{%
 \ifx #1\expandafter \@firstoftwo
 \else \expandafter \@secondoftwo
 \fi
}%
\providecommand \natexlab [1]{#1}%
\providecommand \enquote  [1]{``#1''}%
\providecommand \bibnamefont  [1]{#1}%
\providecommand \bibfnamefont [1]{#1}%
\providecommand \citenamefont [1]{#1}%
\providecommand \href@noop [0]{\@secondoftwo}%
\providecommand \href [0]{\begingroup \@sanitize@url \@href}%
\providecommand \@href[1]{\@@startlink{#1}\@@href}%
\providecommand \@@href[1]{\endgroup#1\@@endlink}%
\providecommand \@sanitize@url [0]{\catcode `\\12\catcode `\$12\catcode `\&12\catcode `\#12\catcode `\^12\catcode `\_12\catcode `\%12\relax}%
\providecommand \@@startlink[1]{}%
\providecommand \@@endlink[0]{}%
\providecommand \url  [0]{\begingroup\@sanitize@url \@url }%
\providecommand \@url [1]{\endgroup\@href {#1}{\urlprefix }}%
\providecommand \urlprefix  [0]{URL }%
\providecommand \Eprint [0]{\href }%
\providecommand \doibase [0]{http://dx.doi.org/}%
\providecommand \selectlanguage [0]{\@gobble}%
\providecommand \bibinfo  [0]{\@secondoftwo}%
\providecommand \bibfield  [0]{\@secondoftwo}%
\providecommand \translation [1]{[#1]}%
\providecommand \BibitemOpen [0]{}%
\providecommand \bibitemStop [0]{}%
\providecommand \bibitemNoStop [0]{.\EOS\space}%
\providecommand \EOS [0]{\spacefactor3000\relax}%
\providecommand \BibitemShut  [1]{\csname bibitem#1\endcsname}%
\let\auto@bib@innerbib\@empty
\bibitem [{\citenamefont {Wang}\ \emph {et~al.}(2020)\citenamefont {Wang}, \citenamefont {Zhang}, \citenamefont {Fan}, \citenamefont {Ren}, \citenamefont {Song}, \citenamefont {Ma},\ and\ \citenamefont {Xue}}]{CDW_SnSe2}%
  \BibitemOpen
  \bibfield  {author} {\bibinfo {author} {\bibfnamefont {S.-Z.}\ \bibnamefont {Wang}}, \bibinfo {author} {\bibfnamefont {Y.-M.}\ \bibnamefont {Zhang}}, \bibinfo {author} {\bibfnamefont {J.-Q.}\ \bibnamefont {Fan}}, \bibinfo {author} {\bibfnamefont {M.-Q.}\ \bibnamefont {Ren}}, \bibinfo {author} {\bibfnamefont {C.-L.}\ \bibnamefont {Song}}, \bibinfo {author} {\bibfnamefont {X.-C.}\ \bibnamefont {Ma}}, \ and\ \bibinfo {author} {\bibfnamefont {Q.-K.}\ \bibnamefont {Xue}},\ }\href@noop {} {\bibfield  {journal} {\bibinfo  {journal} {Phys. Rev. B}\ }\textbf {\bibinfo {volume} {102}},\ \bibinfo {pages} {241408} (\bibinfo {year} {2020})}\BibitemShut {NoStop}%
\bibitem [{\citenamefont {Zhang}\ \emph {et~al.}(2018{\natexlab{a}})\citenamefont {Zhang}, \citenamefont {Fan}, \citenamefont {Wang}, \citenamefont {Zhang}, \citenamefont {Wang}, \citenamefont {Li}, \citenamefont {He}, \citenamefont {Song}, \citenamefont {Ma},\ and\ \citenamefont {Xue}}]{Superconductivty}%
  \BibitemOpen
  \bibfield  {author} {\bibinfo {author} {\bibfnamefont {Y.-M.}\ \bibnamefont {Zhang}}, \bibinfo {author} {\bibfnamefont {J.-Q.}\ \bibnamefont {Fan}}, \bibinfo {author} {\bibfnamefont {W.-L.}\ \bibnamefont {Wang}}, \bibinfo {author} {\bibfnamefont {D.}~\bibnamefont {Zhang}}, \bibinfo {author} {\bibfnamefont {L.}~\bibnamefont {Wang}}, \bibinfo {author} {\bibfnamefont {W.}~\bibnamefont {Li}}, \bibinfo {author} {\bibfnamefont {K.}~\bibnamefont {He}}, \bibinfo {author} {\bibfnamefont {C.-L.}\ \bibnamefont {Song}}, \bibinfo {author} {\bibfnamefont {X.-C.}\ \bibnamefont {Ma}}, \ and\ \bibinfo {author} {\bibfnamefont {Q.-K.}\ \bibnamefont {Xue}},\ }\href@noop {} {\bibfield  {journal} {\bibinfo  {journal} {Phys. Rev. B}\ }\textbf {\bibinfo {volume} {98}},\ \bibinfo {pages} {220508} (\bibinfo {year} {2018}{\natexlab{a}})}\BibitemShut {NoStop}%
\bibitem [{\citenamefont {Ma}\ \emph {et~al.}(2020)\citenamefont {Ma}, \citenamefont {Shi}, \citenamefont {Kang}, \citenamefont {Peng}, \citenamefont {Meng}, \citenamefont {Zhu}, \citenamefont {Cui}, \citenamefont {Sun}, \citenamefont {Ma}, \citenamefont {Wang}, \citenamefont {Lei}, \citenamefont {Wu},\ and\ \citenamefont {Chen}}]{Superconductivty_2}%
  \BibitemOpen
  \bibfield  {author} {\bibinfo {author} {\bibfnamefont {L.~K.}\ \bibnamefont {Ma}}, \bibinfo {author} {\bibfnamefont {M.~Z.}\ \bibnamefont {Shi}}, \bibinfo {author} {\bibfnamefont {B.~L.}\ \bibnamefont {Kang}}, \bibinfo {author} {\bibfnamefont {K.~L.}\ \bibnamefont {Peng}}, \bibinfo {author} {\bibfnamefont {F.~B.}\ \bibnamefont {Meng}}, \bibinfo {author} {\bibfnamefont {C.~S.}\ \bibnamefont {Zhu}}, \bibinfo {author} {\bibfnamefont {J.~H.}\ \bibnamefont {Cui}}, \bibinfo {author} {\bibfnamefont {Z.~L.}\ \bibnamefont {Sun}}, \bibinfo {author} {\bibfnamefont {D.~H.}\ \bibnamefont {Ma}}, \bibinfo {author} {\bibfnamefont {H.~H.}\ \bibnamefont {Wang}}, \bibinfo {author} {\bibfnamefont {B.}~\bibnamefont {Lei}}, \bibinfo {author} {\bibfnamefont {T.}~\bibnamefont {Wu}}, \ and\ \bibinfo {author} {\bibfnamefont {X.~H.}\ \bibnamefont {Chen}},\ }\href@noop {} {\bibfield  {journal} {\bibinfo  {journal} {Phys. Rev. Mater.}\ }\textbf {\bibinfo {volume} {4}},\ \bibinfo {pages} {124803} (\bibinfo {year} {2020})}\BibitemShut
  {NoStop}%
\bibitem [{\citenamefont {Domingo}, \citenamefont {Itoga},\ and\ \citenamefont {Kannewurf}(1966)}]{domingo1966fundamental}%
  \BibitemOpen
  \bibfield  {author} {\bibinfo {author} {\bibfnamefont {G.}~\bibnamefont {Domingo}}, \bibinfo {author} {\bibfnamefont {R.~S.}\ \bibnamefont {Itoga}}, \ and\ \bibinfo {author} {\bibfnamefont {C.~R.}\ \bibnamefont {Kannewurf}},\ }\href@noop {} {\bibfield  {journal} {\bibinfo  {journal} {Phys. Rev.}\ }\textbf {\bibinfo {volume} {143}},\ \bibinfo {pages} {536} (\bibinfo {year} {1966})}\BibitemShut {NoStop}%
\bibitem [{\citenamefont {Pan}\ \emph {et~al.}(2013)\citenamefont {Pan}, \citenamefont {De}, \citenamefont {Manongdo}, \citenamefont {Guloy}, \citenamefont {Hadjiev}, \citenamefont {Lin},\ and\ \citenamefont {Peng}}]{5}%
  \BibitemOpen
  \bibfield  {author} {\bibinfo {author} {\bibfnamefont {T.}~\bibnamefont {Pan}}, \bibinfo {author} {\bibfnamefont {D.}~\bibnamefont {De}}, \bibinfo {author} {\bibfnamefont {J.}~\bibnamefont {Manongdo}}, \bibinfo {author} {\bibfnamefont {A.}~\bibnamefont {Guloy}}, \bibinfo {author} {\bibfnamefont {V.}~\bibnamefont {Hadjiev}}, \bibinfo {author} {\bibfnamefont {Y.}~\bibnamefont {Lin}}, \ and\ \bibinfo {author} {\bibfnamefont {H.}~\bibnamefont {Peng}},\ }\href@noop {} {\bibfield  {journal} {\bibinfo  {journal} {Appl. Phys. Lett.}\ }\textbf {\bibinfo {volume} {103}},\ \bibinfo {pages} {093108} (\bibinfo {year} {2013})}\BibitemShut {NoStop}%
\bibitem [{\citenamefont {Pei}\ \emph {et~al.}(2016)\citenamefont {Pei}, \citenamefont {Bao}, \citenamefont {Wang}, \citenamefont {Ma}, \citenamefont {Yang}, \citenamefont {Li}, \citenamefont {Gu}, \citenamefont {Pantelides}, \citenamefont {Du},\ and\ \citenamefont {Gao}}]{8}%
  \BibitemOpen
  \bibfield  {author} {\bibinfo {author} {\bibfnamefont {T.}~\bibnamefont {Pei}}, \bibinfo {author} {\bibfnamefont {L.}~\bibnamefont {Bao}}, \bibinfo {author} {\bibfnamefont {G.}~\bibnamefont {Wang}}, \bibinfo {author} {\bibfnamefont {R.}~\bibnamefont {Ma}}, \bibinfo {author} {\bibfnamefont {H.}~\bibnamefont {Yang}}, \bibinfo {author} {\bibfnamefont {J.}~\bibnamefont {Li}}, \bibinfo {author} {\bibfnamefont {C.}~\bibnamefont {Gu}}, \bibinfo {author} {\bibfnamefont {S.}~\bibnamefont {Pantelides}}, \bibinfo {author} {\bibfnamefont {S.}~\bibnamefont {Du}}, \ and\ \bibinfo {author} {\bibfnamefont {H.-j.}\ \bibnamefont {Gao}},\ }\href@noop {} {\bibfield  {journal} {\bibinfo  {journal} {Appl. Phys. Lett.}\ }\textbf {\bibinfo {volume} {108}},\ \bibinfo {pages} {053506} (\bibinfo {year} {2016})}\BibitemShut {NoStop}%
\bibitem [{\citenamefont {Su}\ \emph {et~al.}(2013)\citenamefont {Su}, \citenamefont {Ebrish}, \citenamefont {Olson},\ and\ \citenamefont {Koester}}]{6}%
  \BibitemOpen
  \bibfield  {author} {\bibinfo {author} {\bibfnamefont {Y.}~\bibnamefont {Su}}, \bibinfo {author} {\bibfnamefont {M.~A.}\ \bibnamefont {Ebrish}}, \bibinfo {author} {\bibfnamefont {E.~J.}\ \bibnamefont {Olson}}, \ and\ \bibinfo {author} {\bibfnamefont {S.~J.}\ \bibnamefont {Koester}},\ }\href@noop {} {\bibfield  {journal} {\bibinfo  {journal} {Appl. Phys. Lett.}\ }\textbf {\bibinfo {volume} {103}},\ \bibinfo {pages} {263104} (\bibinfo {year} {2013})}\BibitemShut {NoStop}%
\bibitem [{\citenamefont {Abuzaid}, \citenamefont {Williams},\ and\ \citenamefont {Franklin}(2021)}]{7}%
  \BibitemOpen
  \bibfield  {author} {\bibinfo {author} {\bibfnamefont {H.}~\bibnamefont {Abuzaid}}, \bibinfo {author} {\bibfnamefont {N.~X.}\ \bibnamefont {Williams}}, \ and\ \bibinfo {author} {\bibfnamefont {A.~D.}\ \bibnamefont {Franklin}},\ }\href@noop {} {\bibfield  {journal} {\bibinfo  {journal} {Appl. Phys. Lett.}\ }\textbf {\bibinfo {volume} {118}},\ \bibinfo {pages} {030501} (\bibinfo {year} {2021})}\BibitemShut {NoStop}%
\bibitem [{\citenamefont {Guo}\ \emph {et~al.}(2016)\citenamefont {Guo}, \citenamefont {Tian}, \citenamefont {Xiao}, \citenamefont {Mi},\ and\ \citenamefont {Xue}}]{9}%
  \BibitemOpen
  \bibfield  {author} {\bibinfo {author} {\bibfnamefont {C.}~\bibnamefont {Guo}}, \bibinfo {author} {\bibfnamefont {Z.}~\bibnamefont {Tian}}, \bibinfo {author} {\bibfnamefont {Y.}~\bibnamefont {Xiao}}, \bibinfo {author} {\bibfnamefont {Q.}~\bibnamefont {Mi}}, \ and\ \bibinfo {author} {\bibfnamefont {J.}~\bibnamefont {Xue}},\ }\href@noop {} {\bibfield  {journal} {\bibinfo  {journal} {Appl. Phys. Lett.}\ }\textbf {\bibinfo {volume} {109}},\ \bibinfo {pages} {203104} (\bibinfo {year} {2016})}\BibitemShut {NoStop}%
\bibitem [{\citenamefont {Le~Bail}, \citenamefont {Duroy},\ and\ \citenamefont {Fourquet}(1988)}]{lebail}%
  \BibitemOpen
  \bibfield  {author} {\bibinfo {author} {\bibfnamefont {A.}~\bibnamefont {Le~Bail}}, \bibinfo {author} {\bibfnamefont {H.}~\bibnamefont {Duroy}}, \ and\ \bibinfo {author} {\bibfnamefont {J.~L.}\ \bibnamefont {Fourquet}},\ }\href@noop {} {\bibfield  {journal} {\bibinfo  {journal} {Mater. Res. Bull.}\ }\textbf {\bibinfo {volume} {23}},\ \bibinfo {pages} {447} (\bibinfo {year} {1988})}\BibitemShut {NoStop}%
\bibitem [{\citenamefont {Rodr{\'\i}guez-Carvajal}(1993)}]{carvajal}%
  \BibitemOpen
  \bibfield  {author} {\bibinfo {author} {\bibfnamefont {J.}~\bibnamefont {Rodr{\'\i}guez-Carvajal}},\ }\href@noop {} {\bibfield  {journal} {\bibinfo  {journal} {Physica B: Condens. Matter.}\ }\textbf {\bibinfo {volume} {192}},\ \bibinfo {pages} {55} (\bibinfo {year} {1993})}\BibitemShut {NoStop}%
\bibitem [{\citenamefont {Pham}\ \emph {et~al.}(2020)\citenamefont {Pham}, \citenamefont {Vu}, \citenamefont {Cheng}, \citenamefont {Trinh}, \citenamefont {Lee}, \citenamefont {Ryu}, \citenamefont {Hwang}, \citenamefont {Mo}, \citenamefont {Kim}, \citenamefont {Zhao} \emph {et~al.}}]{pham2020high}%
  \BibitemOpen
  \bibfield  {author} {\bibinfo {author} {\bibfnamefont {A.-T.}\ \bibnamefont {Pham}}, \bibinfo {author} {\bibfnamefont {T.~H.}\ \bibnamefont {Vu}}, \bibinfo {author} {\bibfnamefont {C.}~\bibnamefont {Cheng}}, \bibinfo {author} {\bibfnamefont {T.~L.}\ \bibnamefont {Trinh}}, \bibinfo {author} {\bibfnamefont {J.-E.}\ \bibnamefont {Lee}}, \bibinfo {author} {\bibfnamefont {H.}~\bibnamefont {Ryu}}, \bibinfo {author} {\bibfnamefont {C.}~\bibnamefont {Hwang}}, \bibinfo {author} {\bibfnamefont {S.-K.}\ \bibnamefont {Mo}}, \bibinfo {author} {\bibfnamefont {J.}~\bibnamefont {Kim}}, \bibinfo {author} {\bibfnamefont {L.-d.}\ \bibnamefont {Zhao}},  \emph {et~al.},\ }\href@noop {} {\bibfield  {journal} {\bibinfo  {journal} {ACS Appl. Energy Mater.}\ }\textbf {\bibinfo {volume} {3}},\ \bibinfo {pages} {10787} (\bibinfo {year} {2020})}\BibitemShut {NoStop}%
\bibitem [{\citenamefont {Biswas}\ \emph {et~al.}(2021)\citenamefont {Biswas}, \citenamefont {Dandu}, \citenamefont {Prosad}, \citenamefont {Das}, \citenamefont {Menon}, \citenamefont {Deka}, \citenamefont {Majumdar},\ and\ \citenamefont {Raghunathan}}]{13}%
  \BibitemOpen
  \bibfield  {author} {\bibinfo {author} {\bibfnamefont {R.}~\bibnamefont {Biswas}}, \bibinfo {author} {\bibfnamefont {M.}~\bibnamefont {Dandu}}, \bibinfo {author} {\bibfnamefont {A.}~\bibnamefont {Prosad}}, \bibinfo {author} {\bibfnamefont {S.}~\bibnamefont {Das}}, \bibinfo {author} {\bibfnamefont {S.}~\bibnamefont {Menon}}, \bibinfo {author} {\bibfnamefont {J.}~\bibnamefont {Deka}}, \bibinfo {author} {\bibfnamefont {K.}~\bibnamefont {Majumdar}}, \ and\ \bibinfo {author} {\bibfnamefont {V.}~\bibnamefont {Raghunathan}},\ }\href@noop {} {\bibfield  {journal} {\bibinfo  {journal} {Sci. Rep.}\ }\textbf {\bibinfo {volume} {11}},\ \bibinfo {pages} {15017} (\bibinfo {year} {2021})}\BibitemShut {NoStop}%
\bibitem [{\citenamefont {Mandal}\ \emph {et~al.}(2024)\citenamefont {Mandal}, \citenamefont {Maity}, \citenamefont {Barman}, \citenamefont {Biswas}, \citenamefont {Mondal}, \citenamefont {Raghunathan}, \citenamefont {Singh}, \citenamefont {Nayak},\ and\ \citenamefont {Sethupathi}}]{mandal2024low}%
  \BibitemOpen
  \bibfield  {author} {\bibinfo {author} {\bibfnamefont {M.}~\bibnamefont {Mandal}}, \bibinfo {author} {\bibfnamefont {N.}~\bibnamefont {Maity}}, \bibinfo {author} {\bibfnamefont {P.~K.}\ \bibnamefont {Barman}}, \bibinfo {author} {\bibfnamefont {R.}~\bibnamefont {Biswas}}, \bibinfo {author} {\bibfnamefont {S.}~\bibnamefont {Mondal}}, \bibinfo {author} {\bibfnamefont {V.}~\bibnamefont {Raghunathan}}, \bibinfo {author} {\bibfnamefont {A.~K.}\ \bibnamefont {Singh}}, \bibinfo {author} {\bibfnamefont {P.~K.}\ \bibnamefont {Nayak}}, \ and\ \bibinfo {author} {\bibfnamefont {K.}~\bibnamefont {Sethupathi}},\ }\href@noop {} {\bibfield  {journal} {\bibinfo  {journal} {Phys. Rev. B}\ }\textbf {\bibinfo {volume} {110}},\ \bibinfo {pages} {195404} (\bibinfo {year} {2024})}\BibitemShut {NoStop}%
\bibitem [{\citenamefont {Ying}\ \emph {et~al.}(2018)\citenamefont {Ying}, \citenamefont {Paudyal}, \citenamefont {Heil}, \citenamefont {Chen}, \citenamefont {Struzhkin},\ and\ \citenamefont {Margine}}]{ying2018unusual}%
  \BibitemOpen
  \bibfield  {author} {\bibinfo {author} {\bibfnamefont {J.}~\bibnamefont {Ying}}, \bibinfo {author} {\bibfnamefont {H.}~\bibnamefont {Paudyal}}, \bibinfo {author} {\bibfnamefont {C.}~\bibnamefont {Heil}}, \bibinfo {author} {\bibfnamefont {X.-J.}\ \bibnamefont {Chen}}, \bibinfo {author} {\bibfnamefont {V.~V.}\ \bibnamefont {Struzhkin}}, \ and\ \bibinfo {author} {\bibfnamefont {E.~R.}\ \bibnamefont {Margine}},\ }\href@noop {} {\bibfield  {journal} {\bibinfo  {journal} {Phys. Rev. Lett.}\ }\textbf {\bibinfo {volume} {121}},\ \bibinfo {pages} {027003} (\bibinfo {year} {2018})}\BibitemShut {NoStop}%
\bibitem [{\citenamefont {Pallecchi}\ \emph {et~al.}(2023)\citenamefont {Pallecchi}, \citenamefont {Caglieris}, \citenamefont {Ceccardi}, \citenamefont {Manca}, \citenamefont {Marr\'e}, \citenamefont {Repetto}, \citenamefont {Schott}, \citenamefont {Bilc}, \citenamefont {Chaitoglou}, \citenamefont {Dimoulas},\ and\ \citenamefont {Verstraete}}]{19}%
  \BibitemOpen
  \bibfield  {author} {\bibinfo {author} {\bibfnamefont {I.}~\bibnamefont {Pallecchi}}, \bibinfo {author} {\bibfnamefont {F.}~\bibnamefont {Caglieris}}, \bibinfo {author} {\bibfnamefont {M.}~\bibnamefont {Ceccardi}}, \bibinfo {author} {\bibfnamefont {N.}~\bibnamefont {Manca}}, \bibinfo {author} {\bibfnamefont {D.}~\bibnamefont {Marr\'e}}, \bibinfo {author} {\bibfnamefont {L.}~\bibnamefont {Repetto}}, \bibinfo {author} {\bibfnamefont {M.}~\bibnamefont {Schott}}, \bibinfo {author} {\bibfnamefont {D.~I.}\ \bibnamefont {Bilc}}, \bibinfo {author} {\bibfnamefont {S.}~\bibnamefont {Chaitoglou}}, \bibinfo {author} {\bibfnamefont {A.}~\bibnamefont {Dimoulas}}, \ and\ \bibinfo {author} {\bibfnamefont {M.~J.}\ \bibnamefont {Verstraete}},\ }\href@noop {} {\bibfield  {journal} {\bibinfo  {journal} {Phys. Rev. Mater.}\ }\textbf {\bibinfo {volume} {7}},\ \bibinfo {pages} {054004} (\bibinfo {year} {2023})}\BibitemShut {NoStop}%
\bibitem [{\citenamefont {Mott}(1969)}]{mott1969conduction}%
  \BibitemOpen
  \bibfield  {author} {\bibinfo {author} {\bibfnamefont {N.~F.}\ \bibnamefont {Mott}},\ }\href@noop {} {\bibfield  {journal} {\bibinfo  {journal} {Philos. Mag.}\ }\textbf {\bibinfo {volume} {19}},\ \bibinfo {pages} {835} (\bibinfo {year} {1969})}\BibitemShut {NoStop}%
\bibitem [{\citenamefont {Zhang}\ \emph {et~al.}(2018{\natexlab{b}})\citenamefont {Zhang}, \citenamefont {Li}, \citenamefont {Lochocki}, \citenamefont {Vishwanath}, \citenamefont {Liu}, \citenamefont {Yan}, \citenamefont {Lien}, \citenamefont {Dobrowolska}, \citenamefont {Furdyna}, \citenamefont {Shen} \emph {et~al.}}]{highelectron_affinity_SnSe2}%
  \BibitemOpen
  \bibfield  {author} {\bibinfo {author} {\bibfnamefont {Q.}~\bibnamefont {Zhang}}, \bibinfo {author} {\bibfnamefont {M.~O.}\ \bibnamefont {Li}}, \bibinfo {author} {\bibfnamefont {E.~B.}\ \bibnamefont {Lochocki}}, \bibinfo {author} {\bibfnamefont {S.}~\bibnamefont {Vishwanath}}, \bibinfo {author} {\bibfnamefont {X.}~\bibnamefont {Liu}}, \bibinfo {author} {\bibfnamefont {R.}~\bibnamefont {Yan}}, \bibinfo {author} {\bibfnamefont {H.-H.}\ \bibnamefont {Lien}}, \bibinfo {author} {\bibfnamefont {M.}~\bibnamefont {Dobrowolska}}, \bibinfo {author} {\bibfnamefont {J.}~\bibnamefont {Furdyna}}, \bibinfo {author} {\bibfnamefont {K.~M.}\ \bibnamefont {Shen}},  \emph {et~al.},\ }\href@noop {} {\bibfield  {journal} {\bibinfo  {journal} {Appl. Phys. Lett.}\ }\textbf {\bibinfo {volume} {112}} (\bibinfo {year} {2018}{\natexlab{b}})}\BibitemShut {NoStop}%
\bibitem [{\citenamefont {Zhong}\ \emph {et~al.}(2020)\citenamefont {Zhong}, \citenamefont {Yu}, \citenamefont {Kuang}, \citenamefont {Huang},\ and\ \citenamefont {Yuan}}]{zhong2020electronic}%
  \BibitemOpen
  \bibfield  {author} {\bibinfo {author} {\bibfnamefont {H.}~\bibnamefont {Zhong}}, \bibinfo {author} {\bibfnamefont {J.}~\bibnamefont {Yu}}, \bibinfo {author} {\bibfnamefont {X.}~\bibnamefont {Kuang}}, \bibinfo {author} {\bibfnamefont {K.}~\bibnamefont {Huang}}, \ and\ \bibinfo {author} {\bibfnamefont {S.}~\bibnamefont {Yuan}},\ }\href@noop {} {\bibfield  {journal} {\bibinfo  {journal} {Phys.\ Rev.}\ }\textbf {\bibinfo {volume} {101}},\ \bibinfo {pages} {125430} (\bibinfo {year} {2020})}\BibitemShut {NoStop}%
\bibitem [{\citenamefont {Evans}\ and\ \citenamefont {Hazelwood}(1969)}]{evans1969optical}%
  \BibitemOpen
  \bibfield  {author} {\bibinfo {author} {\bibfnamefont {B.}~\bibnamefont {Evans}}\ and\ \bibinfo {author} {\bibfnamefont {R.}~\bibnamefont {Hazelwood}},\ }\href@noop {} {\bibfield  {journal} {\bibinfo  {journal} {J. Phys. D Appl. Phys.}\ }\textbf {\bibinfo {volume} {2}},\ \bibinfo {pages} {1507} (\bibinfo {year} {1969})}\BibitemShut {NoStop}%
\bibitem [{\citenamefont {Radisavljevic}\ and\ \citenamefont {Kis}(2013)}]{18}%
  \BibitemOpen
  \bibfield  {author} {\bibinfo {author} {\bibfnamefont {B.}~\bibnamefont {Radisavljevic}}\ and\ \bibinfo {author} {\bibfnamefont {A.}~\bibnamefont {Kis}},\ }\href@noop {} {\bibfield  {journal} {\bibinfo  {journal} {Nat. Mater.}\ }\textbf {\bibinfo {volume} {12}},\ \bibinfo {pages} {815} (\bibinfo {year} {2013})}\BibitemShut {NoStop}%
\end{thebibliography}%

\end{document}